# A multicenter study on radiomic features from $T_2$-weighted images of a customized MR pelvic phantom setting the basis for robust radiomic models in clinics

## Authors

Linda Bianchini*[1], João Santinha*[2,3], Nuno Loução[4], Mário Figueiredo[3], Francesca Botta[5], Daniela Origgi[5], Marta Cremonesi[6], Enrico Cassano[7], Nikolaos Papanikolaou[2+] and Alessandro Lascialfari[8+].

*shared first co-authors

[+]shared last co-authors

**Contact**: linda.bianchini@unimi.it, joao.santinha@research.fchampalimaud.org

[1]Department of Physics, Università degli Studi di Milano and INSTM RU, Milan, Italy

[2]Computational Clinical Imaging Group, Center for the Unknown (CCU), Champalimaud Foundation, Lisbon, Portugal

[3]Instituto de Telecomunicações, Instituto Superior Técnico, Lisbon, Portugal

[4]Philips Healthcare, Lisbon, Portugal

[5]Medical Physics Unit, IEO, European Institute of Oncology IRCSS, Milan, Italy

[6]Radiation Research Unit, IEO, European Institute of Oncology IRCSS, Milan, Italy

[7]Breast Imaging Division, IEO, European Institute of Oncology IRCSS, Milan, Italy

[8]Department of Physics, Università degli Studi di Pavia and INSTM RU, Pavia, Italy

## Abstract

_Objectives:_ i) To investigate the repeatability and reproducibility of 2D and 3D radiomic features extracted from $T_2$-weighted images of a pelvic *magnetic resonance* (MR) phantom dedicated to radiomic analysis. ii) To share methodologies supporting robust radiomic models in clinical studies based on images acquired with different equipment and acquisition settings.

_Materials and Methods:_

$T_2$-weighted images of a customized pelvic phantom, with non-homogeneous inserts simulating the texture of tumors, were acquired on three scanners of two manufacturers (GE and Philips Healthcare) and two magnetic field strengths (1.5 T and 3 T). Both 2D and 3D $T_2$-weighted sequences were acquired. The repeatability and reproducibility of the radiomic features were assessed respectively by *intraclass correlation coefficient* (ICC) and *concordance correlation coefficient* (CCC), considering repeated acquisitions with or without repositioning the phantom, both under fixed conditions, and with different scanner type, acquisition type, and acquisition parameters. The features showing ICC or CCC > 0.9 were selected, and their dependence on shape information (Spearman correlation coefficient > 0.8) was analyzed. They were also classified for their ability to distinguish different textures, after shuffling the voxel intensities of the original images. The influence of variations in the *time of echo* TE (in the range 80-120 ms) and the *repetition time* TR (4405 and 5000 ms) on the values of the radiomic features was evaluated.

*Results:*

-*Repeatability*: Out of a total of 944, 79.9% to 96.4% of the 2D features across all the scanners showed excellent repeatability in a fixed position. Much lower and wide range (11.2% to 85.4%) was obtained after phantom repositioning. 3D extraction did not improve repeatability performance.

-*Reproducibility on different scanners:* Excellent reproducibility was observed in 4.6% of the features between the two 1.5 T scanners with fixed acquisition parameters (different manufacturers) and in 15.6% of the features, comparing 1.5 T and 3 T (same manufacturer).

-*Combining repeatability and reproducibility:* The set of features showing both excellent repeatability and reproducibility across all the scanners and scenarios corresponds to 3.3% of the total number of features.

-*Reproducibility varying TE or TR:* 82.4% to 94.9% of the features showed excellent agreement when extracted from images acquired with two TEs 5 ms apart. These values decreased when increasing TE intervals. Excellent reproducibility for changes in TR was exhibited by 90.7% of the features.

-*Correlation with shape:* Across all the scanners, 10.6% of the non-shape features were highly correlated with shape and 2.9% to 3.0% of the features showed high correlation between original and intensity-shuffled images. The intersection of these two subsets of features allowed to identify 19 (2.0%) features, which should be excluded *a priori* as they provide only shape - not texture - information.

*Conclusions:* This study demonstrates that radiomic features are affected by specific MRI clinical protocols, even in ideal conditions. The use of our radiomic phantom allowed us to identify drawbacks and unreliable features for radiomic analysis on $T_2$-weighted images, specifically for pelvic imaging. This paper proposes a general workflow to identify repeatable, reproducible, and informative radiomic features, fundamental to ensure robustness of both internal research protocols and multicenter studies.

## 1. Introduction

Computational advances have allowed the development of high-throughput analyses capable of extracting a large number of quantitative features from medical images. This process, called *radiomics* [1]–[3], has been widely studied as a means to provide imaging biomarkers that could assist in patients' disease management, especially in oncology. The predictive models based on radiomic features showed preliminary but promising results in guiding patients' diagnosis, predicting response to treatment and prognosis, and in providing information on cancer genetics [4]. Early clinical demonstrations of radiomics in *magnetic resonance imaging* (MRI) of the pelvis include prostate [5], rectal [6], and cervical [7] cancer, with representative results in pelvic oncology summarized in [8].

Despite the initial promising outcomes, some critical issues of the radiomic methodology process are still unsolved. Radiomic features can be influenced by a plethora of factors [9], such as image acquisition parameters, magnetic field strength, reconstruction algorithms, segmentation intra- and inter-observer variability, image processing, image artifacts, and choice of the software package used for feature extraction. All these factors can hinder the use of radiomic features as imaging biomarkers, requiring repeatability and reproducibility studies to assess their robustness to scanner/acquisition variability (e.g., scanner noise fluctuations, different acquisition parameters, different scanners – possibly with different field strengths), patient variability (anatomical and physiological deviations), and variability in image analysis (e.g., intra- and inter-observer variability when using manual and semi-automatic segmentation) [10]–[12].

The identification of repeatable and reproducible features can serve as an initial feature selection method that ensures the development of reliable biomarkers and models [11]. Some studies have assessed the robustness of radiomic features in different modalities and applications, with the majority of the investigations performed in *computed tomography* (CT) or *positron emission tomography* (PET) [12], [13]. The lack of standardization in the acquisition protocols and signal intensity scales, the high number of interplaying parameters affecting image quality and signal intensity, the lower spatial resolution (compared with CT), and a higher frequency of artifacts make radiomics analysis of MRI data very challenging. Gourtsoyianni et al. [14] studied the repeatability of MRI features extracted from primary rectal cancer. Fiset et al. [15] studied the repeatability and reproducibility of MRI-based radiomic features on cervical cancer. More recently, Schwier and colleagues [12] investigated the repeatability of radiomic features on small prostate tumors, assessing various normalization schemes, image pre-filtering, and bin-widths for image discretization. Despite this thorough inspection, those authors did not find consistent improvements in repeatability across the different approaches. Scalco et al. [16] found that the choice of image intensity normalization technique had a strong impact on the reproducibility of radiomic features, evaluated on $T_2$-weighted ($T_2$-w) MR images of the pelvic region.

Other methodological studies were performed on phantoms, which offer more controlled experimental setups and the possibility of repeated scans. In this regard, most of the existing studies investigated the reliability of radiomic features extracted from CT images. For example, Varghese et al. [17] reported that CT-based texture features depend on the scanner and image acquisition parameters. In another study by Baeßler et al. [18], the repeatability and reproducibility of MRI-based radiomic features under different matrix sizes were investigated for several imaging sequences using fruits and vegetables as test objects. Yang et al. [19] showed that radiomic features depend on the image acquisition process and reconstruction algorithm with experiments on a digital MR phantom. Furthermore, several clinical studies were conducted with variable acquisition parameters, known to affect the tissue contrast, while the influence of such changes in radiomic features was investigated to some extent by Mayerhoefer et al. [20] and Chirra et al. [21].

Although these studies provide some insight into the reliability of radiomic features in MRI, to the best of our knowledge, a systematic investigation on the repeatability and reproducibility of MRI-based radiomic features across clinical scanners/vendors and magnetic field strengths has not yet been performed.

In this study, we used a pelvic phantom designed explicitly for MRI radiomic purposes to assess the repeatability, with and without repositioning, as well as the reproducibility of 2D and 3D radiomic features extracted from $T_2$-w MR images acquired on three different scanners from two vendors and two field strengths. First, we performed the experiments at fixed imaging parameters. Then, we preliminarily investigated the influence of the *repetition time* (TR) and the *time of echo* (TE) variability on the reproducibility of the radiomic features in one of the available scanners.

## 2. Materials and Methods

### 2.1 Study Design

In a clinical study, the database used for a radiomic investigation can include images acquired on different scanners and/or with different imaging protocols. We investigated the robustness of the radiomic features in some of these possible scenarios.

In this multicenter study (schematized in Figure 1), three scanners of two manufactures and two magnetic field strengths were used to assess repeatability and reproducibility of radiomic features. Phantom images were acquired in a test-retest study to investigate the repeatability of the features on each scanner. The acquisitions were performed with a 2D $T_2$-w sequence, optimized on each scanner for pelvic imaging. Between the two repeated scans, the phantom was either kept fixed or repositioned, to evaluate also the repeatability after repositioning. Additionally, repeatability was investigated on 3D features extracted from images acquired using a 3D $T_2$-w sequence on scanner B, to interpret and strengthen the result obtained for the repeatability of features on each scanner for 2D acquisitions.

The reproducibility of the radiomic features was evaluated in three different scenarios: (1) reproducibility at fixed imaging parameters on two scanners of same field strength, but different manufacturers; (2) reproducibility at fixed imaging parameters on two scanners of the same manufacturer, but different field strengths (1.5 T and 3 T); (3) reproducibility with varying TE and TR imaging parameters. Study (1) was performed on scanner A (1.5 T Optima MR450W, General Electric Healthcare, Waukesha, USA) and scanner B (1.5 T Ingenia, Philips Healthcare, Best, the Netherlands). The sequence parameters used for pelvic diagnostic imaging on scanner A were replicated on scanner B, and the phantom was imaged on both scanners. The replication of the same sequence in scanners of different vendors is not exact, as MR sequences are vendor-specific, and not all the parameters are accessible to users. However, considering that some radiomics retrospective studies made use of images acquired on different scanners, we have mimicked a similar situation in ideal conditions (*in phantom* study) to quantify the robustness of features in this scenario and to give useful indications for prospective studies. Study (2) was carried out on scanner B and C (3 T Ingenia, Philips Healthcare, Best, the Netherlands). In this case, it was possible to replicate the sequence parameters on both scanners. Doubling the magnetic field strength, the MR signals – and, thus, the images obtained – are different, as they are influenced by the relaxation times $T_1$ and $T_2$, which depend on the field strength. As a consequence, the radiomic features are expected to vary as well. However, this scenario could be part of a clinical radiomic study and would necessitate the assessment of the reproducibility of features between field strengths. Study (3) was conducted on scanner B (which offered more accessibility due to scheduling and technical reasons), varying TE or TR in the range commonly used when imaging patients for pelvic investigations.

Besides, we evaluated the correlation between texture and shape features, to identify whether the excellent performance in terms of repeatability and reproducibility might be ascribable to high correlation with shape information rather than robust quantification of a texture property.

## 2.2 The Phantom

A pelvis phantom designed in-house for radiomic purposes, PETER PHAN, was used for this study. The procedure to build the phantom was described in detail in [22]. Briefly, a pelvic-shaped container (NEMA IEC Body Phantom Set$^{TM}$, Spectrum Corporation, Durham, USA, https://capintec.com/product/nema-iec-pet-body-phantom-set/) was used as the main phantom compartment, filled with a solution of 0.4 mM $MnCl_2$ to reproduce the relaxation time $T_2$ of the muscle tissue surrounding pelvic lesions. Four inserts (Figure 2a), created by mixing polystyrene spheres of different diameters (1-8 mm) and 0.1 % water solution of agar (Sigma-Aldrich, St. Louis, USA), were embedded in the solution to provide MR average signal and texture consistent with a set of representative pelvic tumors. The use of this phantom has a major advantage of allowing storage and transportation, enabling the study of repeatability and reproducibility over time and of new scanners, which is impossible with other types of phantom based on meat or fruits.

## 2.3 2D Repeatability and Reproducibility between Scanners

### 2.3.1 Images Acquisition and Segmentation

$T_2$-w images of the phantom were acquired on scanners A, B, and C, using different pelvic imaging setups. The parameters of the MR sequences are listed in Table 1. For acquisitions on scanner C (3 T), the aqueous solution in the main phantom compartment was replaced with oil (Spectrasyn 4 phantom oil, Philips Heathcare, Best, the Netherlands) that has a lower dielectric constant than water, thus allowing to avoid specific image artifacts intensified at higher fields.

To discuss the reproducibility results, contrast-to-noise ratio (CNR) was calculated using equation (1) [23], [24], where the mean signal intensity of A, $\mu_{SI_A}$, was measured in the green *region of interest* (ROI) in Figure 3 (corresponding to the insert content), the mean signal intensity of B, $\mu_{SI_B}$, was measured in the yellow ROI (main phantom compartment) and the standard deviation of the signal intensity of background air, $\sigma_{SI_{air}}$, was measured in the red ROI. CNR values are reported in Table 1. Signal-to-noise ratio (SNR) was calculated as well, and details are reported in the Supplementary Materials (sections S.1 and S.2, Table S1).

$$CNR = \frac{\mu_{SI_A} - \mu_{SI_B}}{\sigma_{SI_{air}}} \qquad (1)$$

To test the short-term repeatability of the radiomic features, the acquisition was repeated twice on each scanner without changing the setup nor moving the phantom. Then, the phantom was removed and repositioned, and the acquisition was repeated. Sixteen cylindrical ROIs of four sizes were drawn on the phantom inserts for the first acquisition images (Figure 2b) using 3D Slicer [25] version 4.10.1 (http://www.slicer.org/). Each ROI was drawn on a total of three consecutive slices. The ROIs for the images acquired after phantom repositioning were obtained by applying to the original ROIs segmentation the rigid transformation that allowed to align the initial images with the images of the repositioned phantom.

### 2.3.2 Radiomic Features Extraction

The open-source Python package PyRadiomics [26] version 2.2.0[1] was used to normalize the images and to extract 2D radiomic features (included categories: Shape, First Order, GLCM, GLRLM, GLSZM, NGTDM, and GLDM[2]) from each ROI, on both original and filtered images (Laplacian of Gaussian - LoG -, Wavelet, Square, Square Root, Logarithm, and Exponential). This radiomics software was chosen among others, as it allows the extraction of the majority of the features defined by the IBSI [27] and it is a well-established reference for the development of radiomics. We used a 6.0 mm $\sigma$ in the LoG filter and one level for the wavelet decompositions. PyRadiomics preprocessing normalization was applied by setting the images mean signal intensity to 300 and standard deviation to 100. This normalization was used to reduce the effect of different MR image intensity ranges on different scanners due to the lack of a standard intensity scale in MRI. We chose to use the fixed bin size as the gray-level discretization technique, as suggested by Duron et al. [28], who found a higher number of reproducible radiomic features using this method in contrast to the fixed bin number technique. The bin-width was optimized for each extraction in order to obtain a number of bins in the range 30 to 130, which has shown good performance and reproducibility in the literature [29], as stated in PyRadiomics documentation. In Table S2 in Section S.3 of the Supplementary Material, the main settings used for the feature extraction are listed. The corresponding parameter files are available at https://github.com/ReliabilityRadiomicsIEOFC/PhantomStudy. In Table S3 in Section S.4 of the Supplementary Material, the list of features extracted from each category are reported.

### 2.3.3 Repeatability and Reproducibility Assessment

The repeatability of the radiomic features was quantified by computing the *intraclass correlation coefficient* (ICC) [11], [30], [31]. ICC (2,1), equation (2), was calculated pairwise (between repeated acquisitions) for each radiomic feature to test repeatability, with and without phantom repositioning.

---

[1] available at https://github.com/Radiomics/pyradiomics/tree/2.2.0

[2] GLCM = Gray Level Co-occurrence Matrix; GLRLM = Gray Level Run Length Matrix; GLSZM = Gray Level Size Zone Matrix; NGTDM= Neighbouring Gray Tone Difference Matrix; GLDM= Gray Level Dependence Matrix.

The suffix (2,1) indicates the ICC form computed considering two-way random effects for absolute agreement and single rater/measurement.

$$ICC\ (2{,}1) = \frac{MS_R - MS_E}{MS_R + (k-1)\ MS_E + \frac{k}{n}(MS_C - MS_E)} \qquad (2)$$

In equation (2), $MS_R$ corresponds to the mean square for ROIs, $MS_E$ corresponds to the mean square for error, $MS_C$ corresponds to the mean square for repeated measures, $n$ is the number of ROIs, and $k$ is the number of repeated acquisitions.

To test reproducibility, the *concordance correlation coefficient* (CCC) [32] was calculated pairwise for the features extracted from the images acquired on two different scanners (scanner A vs. B, scanner B vs. C), with the same imaging sequence. In equation (3), $\sigma_1^2$ and $\sigma_2^2$ are the variances of a feature for each acquisition, $\mu_1$and $\mu_2$ are the feature means, and $\rho_{12}$ is the correlation coefficient between the acquisitions.

$$CCC = \frac{2\sigma_1\sigma_2\rho_{12}}{\sigma_1^2 + \sigma_2^2 + (\mu_1 - \mu_2)^2} \qquad (3)$$

The features were classified into four groups based on the ICC or CCC values (Table 2). The most robust features for the particular scenario considered (across three scanners, including two manufacturers and two magnetic field strengths) were identified by intersecting the sets of features showing both excellent repeatability and excellent reproducibility.

### 2.4 Reproducibility under Varying TE or TR

In this experiment, we aimed at investigating features reproducibility under variation of the TE and TR parameters. $T_2$-w images of PETER PHAN were acquired on scanner B using different TE and TR values (Table 1), varied in the range of the usual setting for clinical diagnostic imaging. First, the values of TE were varied between 80 and 120 ms, with a step size of 5 ms for a fixed TR of 5000 ms. Then, to investigate the influence of TR in a $T_2$-w sequence, the Philips scanner mode of TR range (TR: 4000 ms to 6000 ms), which selects the optimal TR within the predefined range, was used to acquire an image with TE of 100 ms and an automatically selected TR of 4405 ms. The phantom inserts were segmented, and the radiomic features extracted as previously described in the other experiments. CCC values were calculated pairwise between the features extracted from images obtained with all possible combinations of TEs and for the images acquired with TR/TE as 5000/100 and 4405/100 to test the features reproducibility.

### 2.5 Repeatability at 3D Level

The repeatability and reproducibility of the radiomic features may be affected by the type of feature extraction performed (2D or 3D level). The majority of clinical MRI acquisitions are 2D, and the downsampling to isotropic voxels - to allow the features extraction in 3D - will likely cause a considerable loss of information in the plane of acquisition. In opposition, the upsampling to isotropic voxel will artificially create new voxel data that may have an impact on the features [27]. Thanks to several technological advances, the acquisition of 3D sequences is becoming a real possibility; therefore, it is interesting to investigate whether 3D features have a better performance in terms of rotation invariance as compared to 2D features, which would result in improved repeatability after phantom repositioning. To investigate this issue, we assessed the rotational invariance of 1316 3D features extracted from images acquired with a 3D $T_2$-w sequence on scanner B with an isotropic voxel 1 mm x 1 mm x 1 mm. The imaging parameters are listed in Table 1. The difference in the number of extracted features between 2D and 3D is a consequence of the increased number of wavelet decompositions in 3D feature extraction. PETER PHAN was scanned with this sequence twice, repositioning the phantom between the acquisitions. The radiomic features were extracted from the 16 ROIs obtained with the procedure explained in the previous experiment. The extraction was performed at 3D level, disabling the "force2D" function on the PyRadiomics parameters file (available at https://github.com/ReliabilityRadiomicsIEOFC/PhantomStudy). The evaluation of the rotational invariance of the features was assessed through the ICC (2,1), calculated pairwise for each feature, comparing its value in the first and the second acquisition.

### 2.6 Assessment of Shape Information in Non-Shape Radiomic Features

Besides repeatability and reproducibility issues, a recent study showed other vulnerabilities of radiomic features by assessing the performance of previously defined models on the same datasets but where images had their voxel intensities randomly shuffled and found that 3 out of 3 non-shape features constituting the model were actually capturing the tumor volume [33].

Following this result, we performed an investigation on the non-shape features, aiming at identifying the radiomic features that were robust because of their high correlation with shape features, but were not extracting the informative content their mathematical definition was expected to quantify. To this purpose, we proceeded with two consecutive analyses.

Firstly, we identified the subset of non-shape features that were highly correlated with shape, through a pairwise-correlation analysis between shape and non-shape radiomic features, evaluated with the Spearman correlation coefficient. Non-shape features showing Spearman's correlation above 0.8 were considered as highly correlated with shape features. Other shape metrics besides volume [34] were included, since characteristics like maximum diameter [35], roundness [36] and spiculation [37], [38], among others, are important oncological diagnostic and prognostic factors.

Secondly, starting from the first acquisitions performed on the three scanners (for scanner B the same sequence used for the repeatability study was considered), three additional sets of images were created by shuffling the voxel intensities of each original image. The radiomic features were extracted from the shuffled intensities datasets, following the procedure described above for the segmentation and the calculation of features. An example of an original image and the corresponding image with randomly shuffled intensities is shown in Figure 4. ICC (2,1) was calculated between the original features and the corresponding features extracted from images that had their intensities randomly shuffled. Features showing ICC > 0.9 were considered texture uninformative since they could not distinguish the original image from the shuffled-intensities one in radiomic terms.

Intersecting the results of the two analyses, it was possible to identify a set of features to exclude *a priori* from an eventual radiomic model since they were both highly correlated with shape and not carrying texture information.

## 3. Results

### 3.1 2D Repeatability and Reproducibility Between Scanners

#### 3.1.1 Repeatability Assessment

A total of 944 features were extracted. Repeatability results for the three scanners, with and without phantom repositioning, are shown in Figure 5. Without repositioning, on scanner A, 910 (96.4%) features showed excellent, 29 (3.1%) good, 3 (0.3%) moderate, and 2 (0.2%) poor repeatability. Similar results were obtained on scanner B, with 869 (92.1%), 58 (6.1%), 8 (0.8%), and 9 (1.0%) features showing respectively excellent, good, moderate, and poor repeatability. The features extracted from the images acquired on scanner C showed less repeatability, with the percentage of features with excellent ICC decreasing to 754 (79.9%), and the percentage of features showing ICC $\leq$ 0.9 increasing to 190 (20.1%). When considering phantom repositioning, a consistent reduction of repeatability was evident across all the scanners. The numbers were 740 (78.4%), 138 (14.6%), 49 (5.2%), 17 (1.8%) (scanner A), 806 (85.4%), 76 (8.1%), 42 (4.4%), 20 (2.1%) (scanner B) and 106 (11.2%), 199 (21.1%), 215 (22.8%), 462 (44.9%) (scanner C) features showing excellent, good, moderate, and poor repeatability.

#### 3.1.2 Reproducibility Assessment

Reproducibility was assessed both in terms of variations between the features extracted from scanners of different manufacturers and equal magnetic field strength and acquisition parameters (A vs. B, Figure 6) and between features extracted from scanners of the same manufacturer but different magnetic field strengths (B vs. C, Figure 6). The CCC values of A vs. B and B vs. C, and corresponding confidence intervals are available at https://github.com/ReliabilityRadiomicsIEOFC/PhantomStudy.

The analysis of the features extracted from the images acquired on scanners A and B showed that 830 (87.9%) features had poor reproducibility. Only 43 (4.6%) features exhibited excellent reproducibility; 29 (3.1%) showed good, and 42 (4.4%) moderate reproducibility.

In terms of reproducibility of the features extracted from images obtained with the same sequence parameters on scanners B (1.5 T) and C (3 T), 147 (15.6%), 295 (31.2%), 321 (34.0%), and 181 (19.2%) of them exhibited excellent, good, moderate, and poor reproducibility.

#### 3.1.3 Overall Repeatability and Reproducibility

In order to identify the most stable radiomic features in this study, the ones showing both excellent repeatability and reproducibility were selected.

A total of 31 (3.3%) features showed excellent repeatability (ICC > 0.9) and reproducibility (CCC > 0.9) across all different scenarios studied (repeatability: with and without phantom repositioning; reproducibility: across manufacturers and field strengths). Apart from shape features, expected to be independent of the experiment settings[3], the 20 (2.1%) remaining features are listed in Table 3.

### 3.2 Reproducibility under Varying TE or TR

Figure 7 shows the percentage of radiomic features presenting (A) excellent, (B) good, (C) moderate, and (D) poor reproducibility between different TE values (according to the CCC thresholds in Table 2). Focusing on Figure 7(A), 82.4% to 94.9% of features showed excellent absolute agreement between TEs 5 ms apart for TEs ranging between 80-120 ms (first diagonal on the left). Increasing the TE interval between two experiments, the percentage of features showing excellent reproducibility decreased progressively. When considering a TE interval of 40 ms, the reproducible features were 19.8% of the total.

When changing from a TR of 5000 ms to a TR of 4404 ms, out of the 944 features, 856 (90.7%) showed excellent, 64 (6.8%) good, 22 (2.3%) moderate, and 2 (0.2%) poor reproducibility.

### 3.3 Repeatability at 3D Level

Out of the 1316 3D features, 446 (33.9%) showed excellent, 394 (29.9%) good, 359 (27.3%) moderate, and 117 (8.9%) poor repeatability. Although we cannot directly compare with the 2D repeatable features, we observe that only a relatively small number of 3D features showed excellent repeatability after repositioning and, therefore, rotational invariance. The 13 features showing rotational invariance in more than 80% of the image types (original or filtered images) are presented in Table 4. Of these, 2 features were First Order, 4 GLCM-based, 2 GLRLM-based, 2 GLSZM-based, 1 NGTDM-based, and 2 GLDM-based.

[3] The reproducibility of the shape features shape_Flatness, shape_LeastAxisLength, and shape_Sphericity being 3D features, can be affected by differences in the through-place spacing.

### 3.4 Assessment of Shape Information in Non-Shape Radiomic Features

In this analysis, two subsets of radiomic features were identified: (i) the subset of non-shape features that were highly correlated with shape; (ii) the subset of features showing a high correlation between their value when extracted from the original image and when extracted from the shuffled-intensities image.

In set (i), out of 930 features, 155 (16.7%), 144 (15.5%), and 158 (17.0%) non-shape features were highly correlated with shape features in scanners A, B, and C, respectively. The discrepancy across scanners in non-shape features showing dependence with shape may be due to differences in contrast between scanners. A matrix showing the correlation between the shape and non-shape features is reported, as an example, in Figure S1 in section S.5 of the Supplementary Materials, where it is possible to observe high correlation of non-shape features with several shape features besides volume. 99 (10.6%) features were shared among all scanners. The subset (ii) was made up of 27 (2.9%), 28 (3.0%), and 27 (2.9%) features for scanners A, B, and C, respectively.

When intersecting the features in subsets (i) and (ii) common to all the scanners, 19 (2.0%) features were obtained. In this way, we identified the set of features nominally belonging to texture but providing only shape information, listed in Table 5.

## 4. Discussion

Clinically meaningful radiomic signatures, providing not only high performance, but also good generalizability, should be constructed using high-quality, reliable features. Therefore, it is essential to understand and incorporate the factors affecting the reliability of such features. Our study focused on several aspects that may influence the values of the radiomic features, as was designed to provide information on: repeatability; reproducibility at fixed imaging parameters; shape information in non-shape features; influence of acquisition parameters (TE and TR).

All imaging data, radiomic files, parameter files for feature extraction, and analysis code are provided, so that researchers can extend this analysis and tune these experiments to their needs.

We investigated two types of 2D repeatability, offering measures on scanner-induced variations (without phantom repositioning) and repositioning-induced variations across different scanners. Without repositioning, it is possible to observe that 92.1% to 96.4% of the features showed excellent repeatability in both 1.5 T scanners. The percentage decreased to 79.9% when considering the features extracted from the images acquired at 3 T. The decrease in repeatability at 3 T may be due to the artifacts more frequently affecting images acquired at a higher field strength. Instead, when considering repositioning, a reduction in the number of features with excellent repeatability was observed: 78.4% of features showing excellent repeatability on scanner A, 85.4% on scanner B, and 11.2% on scanner C. Possible causes for the relatively small reduction observed with scanners A and B may be the rotational invariance of some features and some degree of misalignment of the acquisition after repositioning. As for the larger reduction observed with scanner C, after visual inspection of the images, it was possible to observe that the chemical shift artifact was considerably larger than with scanners A and B, as expected, being proportional to the magnetic field strength. Additionally, as chemical shift artifacts occur in the frequency encoding direction due to the coexistence of water and lipid protons in a voxel, a slight change in the positioning of the phantom will be translated into a change in these artifacts.

We also assessed repeatability in 3D MRI acquisition and observed that only 33.9% of 3D radiomic features showed excellent repeatability after repositioning. As these features were extracted from isotropic images, this represents an unexpected result. In fact, the percentage of 2D features showing excellent repeatability with phantom repositioning for scanner B was much higher (85.4%). This seems to suggest that a 3D extraction from isotropic voxels does not increase the repeatability performance if compared with a 2D extraction. A possible explanation may be that the 3D feature extraction on 3D acquisitions, by being able to offer more detail in the through-plane, may result in more unstable features, as the planning after repositioning does not ensure perfect alignment between acquisitions. In fact, a coarser spacing may not affect the performance of radiomic models and may even improve them as features become less susceptible to noise, repositioning, and other artifacts. These considerations justify the stability observed on the 2D features extracted from the 2D acquisitions.

The reproducibility of the features was assessed by comparing radiomic features extracted from images acquired on scanners of different manufacturers with equal magnetic field strength and acquisition parameters, and scanners of the same manufacturer and different magnetic field strengths. In the first case, we observed that only 4.6% of the features show excellent reproducibility. A similar trend was found in the second case, where 15.6% of the features exhibited excellent reproducibility, and most of the features, 53.2%, had moderated or poor reproducibility. Our results showed a higher reproducibility when comparing two scanners of field strength 1.5 T and 3 T from the same manufacturer (even though the relaxation times $T_1$ and $T_2$ affecting the MR signal are dependent on the field strength), than comparing two scanners of 1.5 T with same acquisition parameters but from different vendors. Regarding reproducibility between different vendors at fixed imaging parameters, we observed that the images acquired on the two scanners exhibited different CNR values. Despite the basic principles of the sequence being the same, these differences can be caused by various sources, that cannot be controlled by the user. A first factor is the type of systems, including the digital versus analog and the location of the analog-to-digital converter. As a matter of fact, Philips converts the signal to digital directly on the coil while GE makes it on the magnet. Other factors consist of different preparation and calibration phases, including distinct power optimizations, frequency determination, shimming, and coils tune.

Despite not being focused on radiomic features, previous studies on the inter-scanner and intra-scanner variability, considering both 1.5 T and 3 T images, on a set of organ-specific measures [39] and proton density fat function measurements [40] showed that disagreement could be seen when comparing scanners of different vendors. Given this, our study allowed us to highlight that even in the presence of unavoidable differences between images acquired with two scanners of different manufacturers (quantified for example by CNR and SNR), a subset of features turned out to be reproducible.

In this illustrative study, we considered a controlled pelvic imaging scenario, resulting in only 31 radiomic features (3.3% of the total number of features extracted) showing excellent robustness in the two repeatability settings and the two reproducibility settings. The scenarios of clinical studies may be different from the one herein considered, but an analogous procedure could be applied, focusing on the phantom inserts that better simulate the texture properties of the tissue under investigation.

As demonstrated in the study by Welch and colleagues [33], some features may exhibit dependencies on volume. However, these dependencies may be extended to other shape features that may also contain diagnostic and prognostic information. Therefore, it is crucial to understand if non-shape features may be repeatable and reproducible due to shape information they may contain. A total of 19 non-shape features common to all three scanners appeared to contain only shape information or were heavily dependent on shape. These features showed both a high correlation with shape features and excellent agreement between features extracted from the original images and images with randomly shuffled intensities.

All the results discussed so far used images acquired with fixed MR sequence parameters. However, the need for a large quantity of data to implement radiomic models may lead to the creation of inhomogeneous databases as used in several MRI clinical radiomic studies. This inhomogeneity, also due to the optimization of imaging parameters on each scanner, might affect the values of radiomic features and, consequently, the performance of models [41], [42].

Our preliminary investigation on the influence of different TE and TR within the same study demonstrated that the majority of the features showed excellent reproducibility when the difference in TE was 5 ms. However, the reproducibility decreased as the TE interval increased, as seen in Figure 7. This corroborated our expectations for a $T_2$-w MRI sequence, as a different TE changes the weighting in $T_2$ and CNR, which will lead to variations of $T_2$-w signal intensity depending on the underlying tissue and pathophysiology. As texture depends on contrast, but since TE-induced differences are non-linear, the dependence on contrast is not removed during image normalization. Similarly, when assessing changes in TR, it was observed that 90.7% of the features showed excellent reproducibility between acquisitions with a TR of 5000 ms and with a TR of 4405 ms.

These results suggest that radiomic studies should be conducted with standardized imaging protocols or making use of the features showing excellent reproducibility in the interval of TEs and TRs used, ensuring this way that the observed results are not associated with differences in the acquisition parameters. Following this consideration, multicenter retrospective studies (involving different scanners with variable protocols) should be coupled with methodological studies to understand the relationship between the variability range of the radiomic features and the variation of each sequence parameter separately. Moreover, given the differences observed between Philips and GE technologies, the investigation of the TE/TR impact on scanners from different vendors will be considered as a natural extension of the present study in future developments.

A few previous studies have assessed the stability of radiomic features in different MRI settings. Mayerhoefer and colleagues [20] investigated the sensitivity of texture features to acquisition parameters (including TE ranging from 20 to 125 ms and TR in the range 900-4500 ms) on $T_2$-w images acquired on a 3 T scanner. They found that the imaging parameters influenced the values of features, with this influence increasing with spatial resolution. Although we did not compare different spatial resolutions, our study performed on a 1.5 T scanner extends the findings of those authors to a lower field strength, while making use of a phantom design for patient imaging, as we proved that the texture features values are dependent on the TE and TR parameters.

In another study by Chirra et al. [21], prostate $T_2$-w MRI images of 147 patients from four different sites were used to assess cross-site reproducibility by performing multivariate cross-validation and assessing preparation-induced instability. The authors found that most of the Haralick features were reproducible in over 99% of all cross-site comparisons. However, that study uses different patient populations on each site and assesses non-tumoral regions under the assumption that these should have a similar texture. Besides, part of their pre-processing involves image upsampling, in some cases by a factor of ~11 times in the image through-plane. Such choices make the comparison with our results difficult.

Our study has the following limitations. Firstly, the use of a phantom cannot include all the effects that exist in real clinical scenarios, such as patient motion, rectal/bladder filling, peristalsis, breathing, tissue diffusion and perfusion, intra-patient tissue variability.

Undeniably, a phantom, as such, cannot be exhaustively representative of all tumors (with substantial differences in terms of dimension, internal structure, texture, shape, aggressiveness, etc.), so it cannot identify all the trustable and robust features for patient image analysis. On the other hand, contrarily to the clinical reality, the use of a phantom allows repeating as many acquisitions as desired, to compare results, and to assess the influence of many parameters. In particular, the use of our inhomogeneous phantom allowed the implementation of a procedure for the identification of features that may not be trustworthy and robust. Furthermore, the use of a phantom, in which biological processes are not present, allows the assessment of the reproducibility issues caused by different system types, e.g., digital vs. analog, coils used, and other sequence parameters, which are translated into distinct image properties like the contrast-to-noise ratio that may have an impact on the values of radiomic features as well. Having this in mind, the experiments conducted in this study provide excellent baseline assessments, under very controlled environments, of the stability of features, and allow avoiding rough misleading results that could be derived without an acquainted input selection before the statistical analysis.

Another limitation of our study was the use of data acquired on a restricted number of scanners (two field strengths, two manufacturers) and the focus on the investigation of the radiomics stability on $T_2$-w MRI images, as part of the clinical diagnostic protocol for pelvic imaging. Conversely, this can represent the starting point for further extension. Thus, the imaging data, parameter files, and analysis scripts are made available in order to allow other researchers to perform similar analyses and incorporate new experimental paths. Following these results, it might be interesting to make use of the phantom in an extended multicenter study using different clinically optimized sequences, for a more comprehensive investigation.

In conclusion, this study investigated the robustness of 2D and 3D radiomic features extracted from $T_2$-w images of a pelvic phantom created for MR radiomic analyses. Our methodological investigation quantified the stability and quality of radiomic features in different MRI settings, enlightening important issues towards robust and reliable models.

Based on a workflow designed to test repeatability and reproducibility, features showing the highest performance were identified. Importantly, many of these repeatable and reproducible features turned out to be inadequate for radiomic analysis, e.g., being non-informative or affected by the image acquisition process.  We recommend to apply this, or similar, procedure to strengthen and support each clinical radiomic study, and we invite researchers to join our effort for a more comprehensive multicenter study.

## Acknowledgements

The authors thank the Nuclear Medicine Departments at Champalimaud Clinical Center – Champalimaud Foundation and at European Institute of Oncology for the availability to use the NEMA IEC Body Phantom Set™.

## References

[1] P. Lambin *et al.*, “Radiomics: Extracting more information from medical images using advanced feature analysis,” *Eur. J. Cancer*, vol. 48, no. 4, pp. 441–446, 2012.

[2] H. J. W. L. Aerts *et al.*, “Decoding tumour phenotype by noninvasive imaging using a quantitative radiomics approach,” *Nat. Commun.*, vol. 5, 2014.

[3] R. J. Gillies, P. E. Kinahan, and H. Hricak, “Radiomics: Images Are More than Pictures, They Are Data,” *Radiology*, vol. 278, no. 2, pp. 563–577, 2016.

[4] S. S. F. Yip and H. J. W. L. Aerts, “Applications and limitations of radiomics,” *Phys. Med. Biol.*, vol. 61, no. 13, pp. R150–R166, 2016.

[5] Y. Sun *et al.*, “Multiparametric MRI and radiomics in prostate cancer: a review,” *Australas. Phys. Eng. Sci. Med.*, vol. 42, no. 1, pp. 3–25, 2019.

[6] X. Zhou *et al.*, “Radiomics-Based Pretherapeutic Prediction of Non-response to Neoadjuvant Therapy in Locally Advanced Rectal Cancer,” *Ann. Surg. Oncol.*, 2019.

[7] Z. Li *et al.*, “MR-Based Radiomics Nomogram of Cervical Cancer in Prediction of the Lymph-Vascular Space Invasion preoperatively,” *J. Magn. Reson. Imaging*, pp. 1–7, 2018.

[8] U. Schick *et al.*, “MRI-derived radiomics: Methodology and clinical applications in the field of pelvic oncology,” *Br. J. Radiol.*, vol. 92, no. 1104, 2019.

[9] A. Traverso, L. Wee, A. Dekker, and R. Gillies, “Repeatability and Reproducibility of Radiomic Features: A Systematic Review,” *Int. J. Radiat. Oncol. Biol. Phys.*, vol. 102, no. 4, pp. 1143–1158, 2018.

[10] R. G. Abramson *et al.*, “MRI Biomarkers in Oncology Clinical Trials,” *Magn. Reson. Imaging Clin.*, vol. 24, no. 1, pp. 11–29, 2016.

[11] D. L. Raunig *et al.*, “Quantitative Imaging Biomarkers: A Review of Statistical Methods for Technical Performance Assessment,” *Stat. Methods Med. Res.*, vol. 24, no. 1, pp. 26–27, 2015.

[12] M. Schwier *et al.*, “Repeatability of Multiparametric Prostate MRI Radiomics Features,” *Sci. Rep.*, vol. 9, no. 1, pp. 1–16, 2019.

[13] D. Mackin *et al.*, “Measuring computed tomography scanner variability of radiomics features,” *Invest. Radiol.*, vol. 50, no. 11, pp. 757–765, 2015.

[14] S. Gourtsoyianni *et al.*, “Primary Rectal Cancer : Repeatability of Global and Local-Regional MR Imaging Texture,” *Radiology*, vol. 000, no. 2, pp. 1–10, 2017.

[15] S. Fiset *et al.*, “Repeatability and reproducibility of MRI-based radiomic features in cervical cancer,” *Radiother. Oncol.*, vol. 135, pp. 107–114, 2019.

[16] E. Scalco *et al.*, “T2w-MRI signal normalization affects radiomics features reproducibility,” *Med. Phys.*, 2020.

[17] B. A. Varghese *et al.*, “Reliability of CT-based texture features: Phantom study,” *J. Appl. Clin. Med. Phys.*, vol. 20, no. 8, pp. 155–163, 2019.

[18] B. Baeßler, K. Weiss, and D. Pinto dos Santos, “Robustness and Reproducibility of Radiomics in Magnetic Resonance Imaging,” *Invest. Radiol.*, vol. 54, no. 4, pp. 221–228, Apr. 2019.

[19] F. Yang, N. Dogan, R. Stoyanova, and J. C. Ford, “Evaluation of radiomic texture feature error due to MRI acquisition and reconstruction: A simulation study utilizing ground truth,” *Phys. Medica*, vol. 50, no. December 2017, pp. 26–36, 2018.

[20] M. E. Mayerhoefer, P. Szomolanyi, D. Jirak, A. Materka, and S. Trattnig, “Effects of MRI acquisition parameter variations and protocol heterogeneity on the results of texture analysis and pattern discrimination: An application-oriented study,” *Med. Phys.*, vol. 36, no.

4, pp. 1236–1243, 2009.

[21] P. Chirra *et al.*, “Multisite evaluation of radiomic feature reproducibility and discriminability for identifying peripheral zone prostate tumors on MRI,” *J. Med. Imaging*, vol. 6, no. 02, p. 1, 2019.

[22] L. Bianchini *et al.*, “PETER PHAN : An MRI phantom for the optimisation of radiomic studies of the female pelvis,” *Phys. Medica*, vol. 71, no. C, pp. 71–81, 2020.

[23] S. Yu, G. Dai, Z. Wang, L. Li, X. Wei, and Y. Xie, “A consistency evaluation of signal-to-noise ratio in the quality assessment of human brain magnetic resonance images,” *BMC Med. Imaging*, vol. 18, no. 1, pp. 1–9, 2018.

[24] S. Foxley *et al.*, “High spectral and spatial resolution MRI of age-related changes in murine prostate,” *Magn. Reson. Med.*, vol. 60, no. 3, pp. 575–581, Sep. 2008.

[25] A. Fedorov *et al.*, “3D Slicer as an Image Computing Platform for the Quantitative Imaging Network,” *Magn. Reson. Imaging*, vol. 30, no. 9, pp. 1323–1341, 2012.

[26] J. J. Van Griethuysen *et al.*, “Computational radiomics system to decode the radiographic phenotype,” *Cancer Res.*, vol. 77, no. 21, pp. e104–e107, 2017.

[27] A. Zwanenburg, S. Leger, M. Vallières, and S. Löck, “Image biomarker standardisation initiative,” *arXiv:1612.07003.*, 2016.

[28] L. Duron *et al.*, “Gray-level discretization impacts reproducible MRI radiomics texture features,” *PLoS One*, vol. 14, no. 3, pp. 1–14, 2019.

[29] F. Tixier *et al.*, “Intratumor heterogeneity characterized by textural features on baseline 18F-FDG PET images predicts response to concomitant radiochemotherapy in esophageal cancer.,” *J. Nucl. Med.*, vol. 52, no. 3, pp. 369–78, 2011.

[30] T. K. Koo and M. Y. Li, “A Guideline of Selecting and Reporting Intraclass Correlation Coefficients for Reliability Research.,” *J. Chiropr. Med.*, vol. 15, no. 2, pp. 155–63, 2016.

[31] D. C. Sullivan *et al.*, “Metrology standards for quantitative imaging biomarkers,” *Radiology*, vol. 277, no. 3, pp. 813–825, 2015.

[32] L. I. Lin, “A Concordance Correlation Coefficient to Evaluate Reproducibility,” *Biometrics*, vol. 45, no. 1, pp. 255–268, 1989.

[33] M. L. Welch *et al.*, “Vulnerabilities of radiomic signature development: The need for safeguards,” *Radiother. Oncol.*, vol. 130, pp. 2–9, 2019.

[34] J. Zhang *et al.*, “Relationship between tumor size and survival in Non-Small-Cell Lung Cancer (NSCLC): An analysis of the Surveillance, Epidemiology, and End Results (SEER) registry,” *J. Thorac. Oncol.*, vol. 10, no. 4, pp. 682–690, 2015.

[35] H. Fukuhara *et al.*, “Maximum tumor diameter: A simple independent predictor for biochemical recurrence after radical prostatectomy,” *Prostate Cancer Prostatic Dis.*, vol. 13, no. 3, pp. 244–247, 2010.

[36] S. I. Lee, E. Oliva, P. F. Hahn, and A. H. Russell, “Malignant tumors of the female pelvic floor: Imaging features that determine therapy: Pictorial review,” *Am. J. Roentgenol.*, vol. 196, no. 3 SUPPL., pp. 15–23, 2011.

[37] A. Snoeckx *et al.*, “Evaluation of the solitary pulmonary nodule: size matters, but do not ignore the power of morphology,” *Insights Imaging*, vol. 9, no. 1, pp. 73–86, 2018.

[38] S. Liu, X.-D. Wu, W.-J. Xu, Q. Lin, X.-J. Liu, and Y. Li, “Is There a Correlation between the Presence of a Spiculated Mass on Mammogram and Luminal A Subtype Breast Cancer?,” *Korean J. Radiol.*, vol. 17, no. 6, p. 846, 2016.

[39] C. L. Schlett *et al.*, “Quantitative, Organ-Specific Interscanner and Intrascanner Variability for 3 T Whole-Body Magnetic Resonance Imaging in a Multicenter, Multivendor Study,”

*Invest. Radiol.*, vol. 51, no. 4, pp. 255–265, 2016.

[40] J. K. Jang *et al.*, “Agreement and Reproducibility of Proton Density Fat Fraction Measurements Using Commercial MR Sequences Across Different Platforms: A Multivendor, Multi-Institutional Phantom Experiment,” *Invest. Radiol.*, vol. 54, no. 8, pp. 517–523, 2019.

[41] S. Rizzo *et al.*, “Radiomics: the facts and the challenges of image analysis,” *Eur. Radiol. Exp.*, vol. 2, no. 1, p. 36, 2018.

[42] N. Papanikolaou, C. Matos, and D. M. Koh, “How to develop a meaningful radiomic signature for clinical use in oncologic patients,” *Cancer Imaging*, vol. 20, no. 1, p. 33, Dec. 2020.

## List of Figures

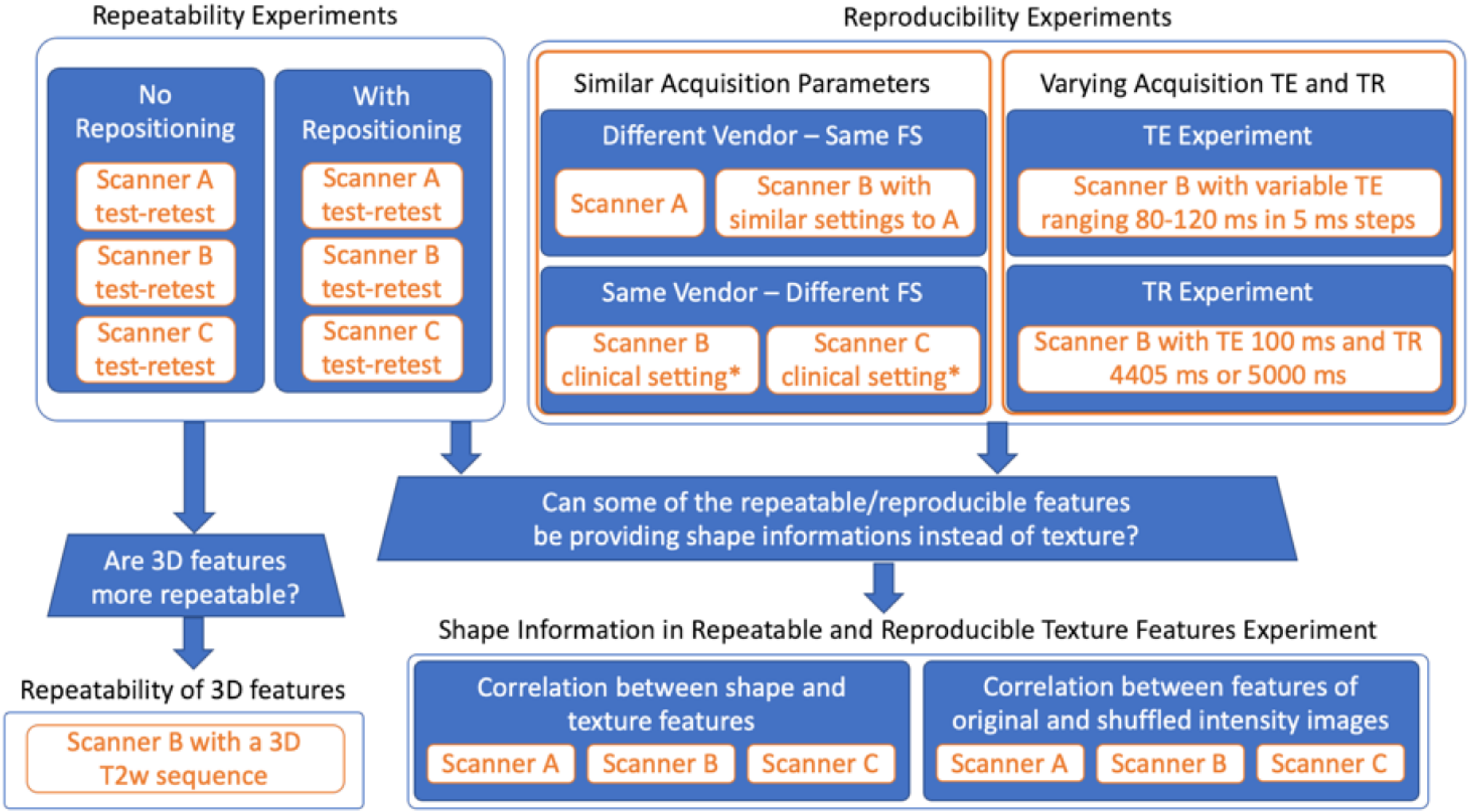

Figure 1. Study design schematic showing repeatability and reproducibility experiments and additional experiments executed to bring understanding to the repeatability and reproducibility results. * - clinical $T_2$-w MRI sequence was shared between scanner B and C. FS refers to the magnetic Field Strength.

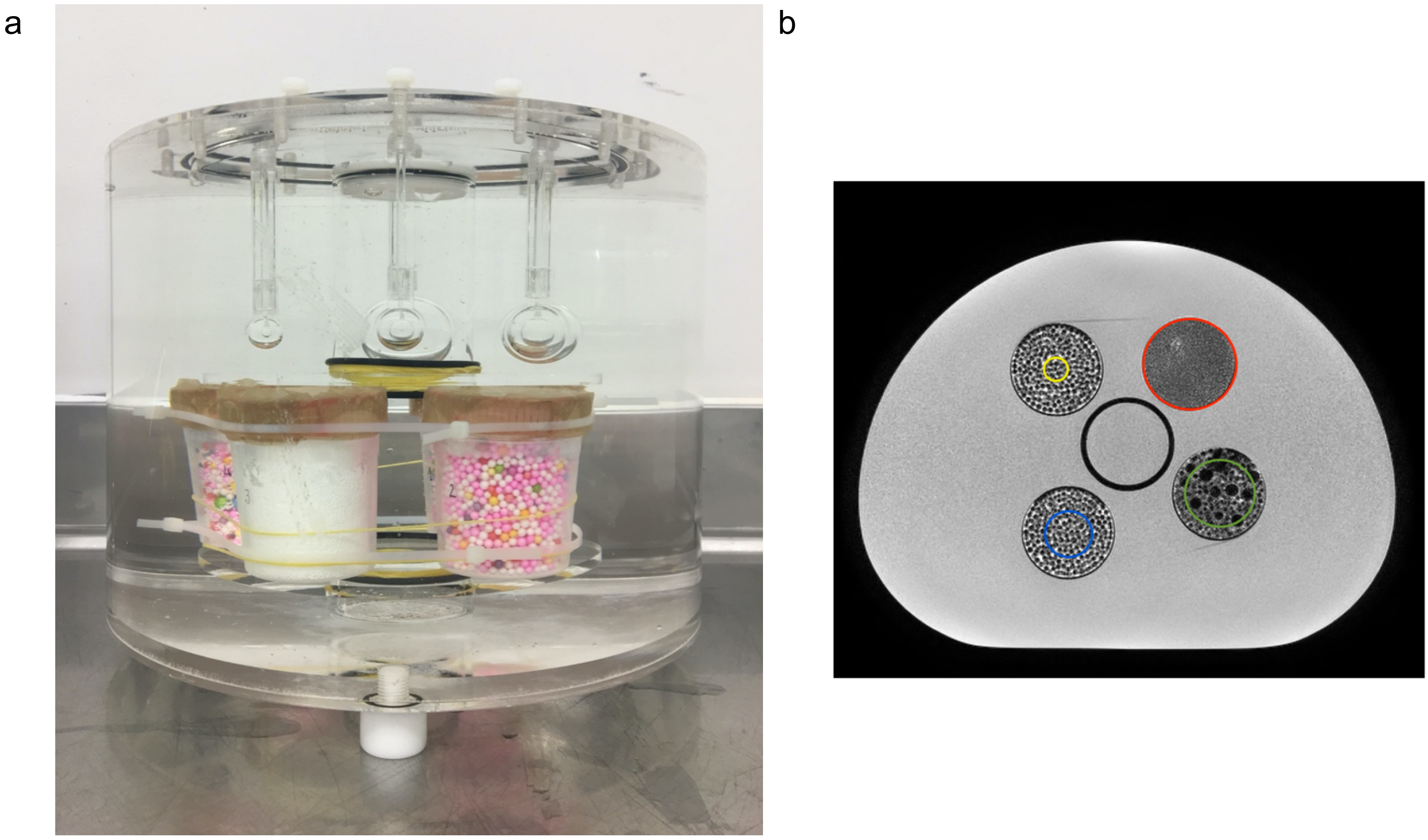

Figure 2. PETER PHAN. (a) Picture of the assembled phantom. The four inserts were filled with spheres of various diameters. One insert contained only small (1 mm) spheres, the second and third inserts were filled with the medium-sized (3-4 mm) spheres and the last one was prepared with a mixture of spheres of small, medium and large (7-8 mm) diameter spheres. (b) Axial $T_2$-weighted image acquired on scanner A with selected ROIs (yellow ∅ = 12 mm; blue ∅ = 24 mm; green ∅ = 36 mm, red ∅ = 48 mm).

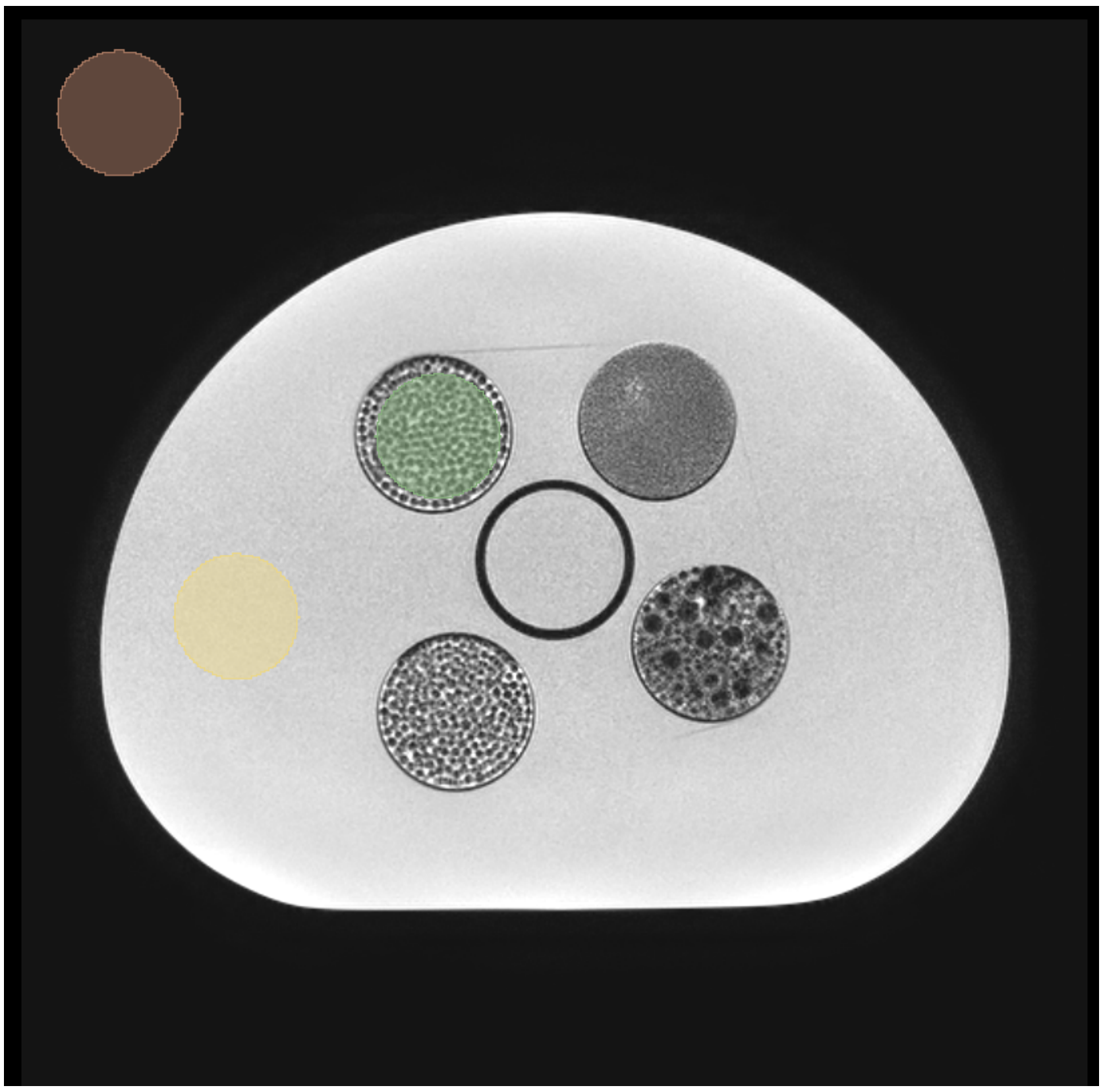

Figure 3. Representation of ROIs placement for calculation of contrast-to-noise ratio and signal-to-noise ratio. Green ROI was used to measure the mean signal intensity of region A, the yellow ROI was used to measure the mean signal intensity of region B, and the red ROI was used to determine the standard deviation of the signal intensity of background air region.

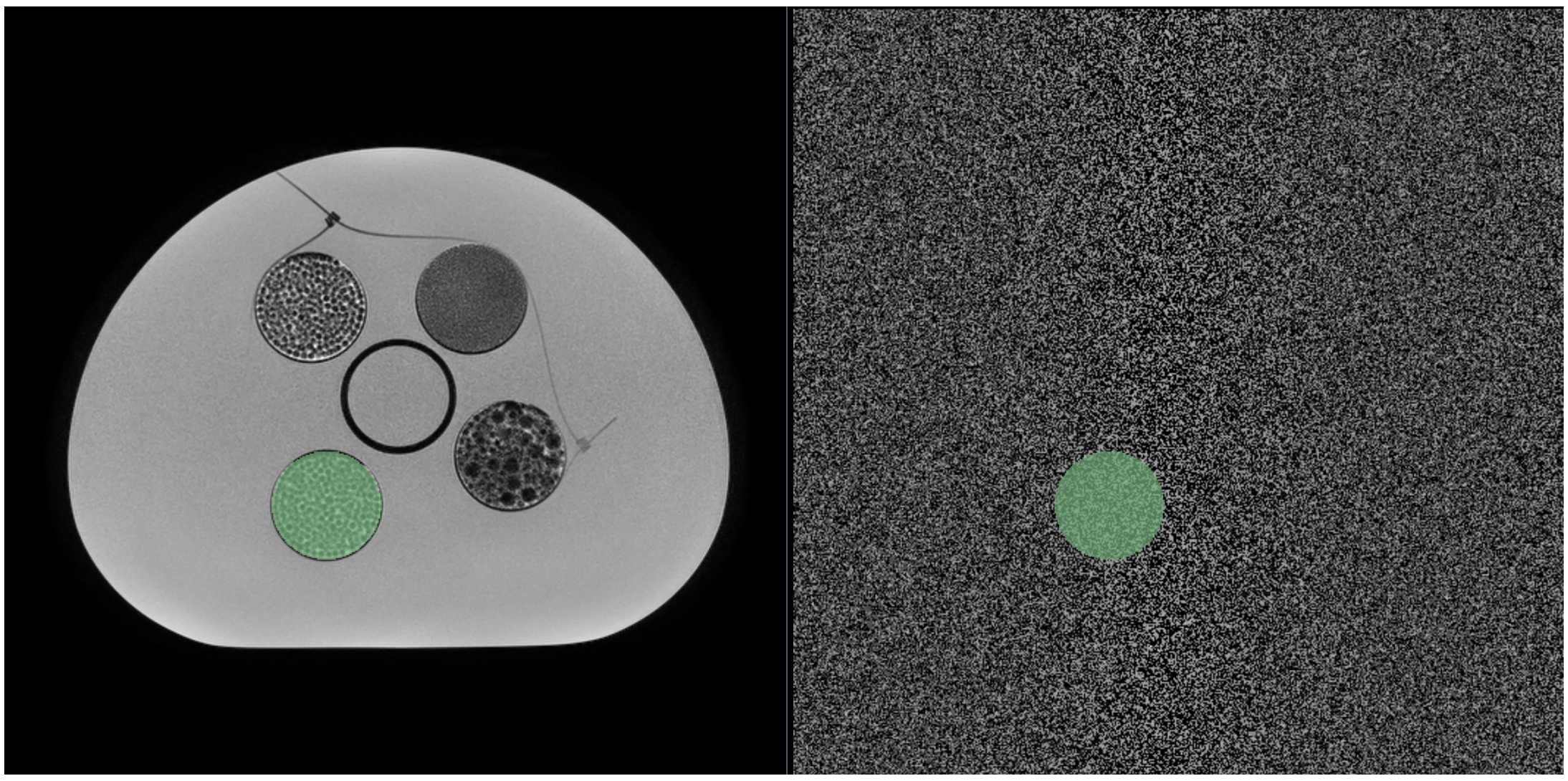

Figure 4. Example of original axial image (left) and of corresponding randomly shuffled intensities axial image (right). The green overlay exemplifies one of the 16 cylindrical regions of interest delineated for the extraction of features.

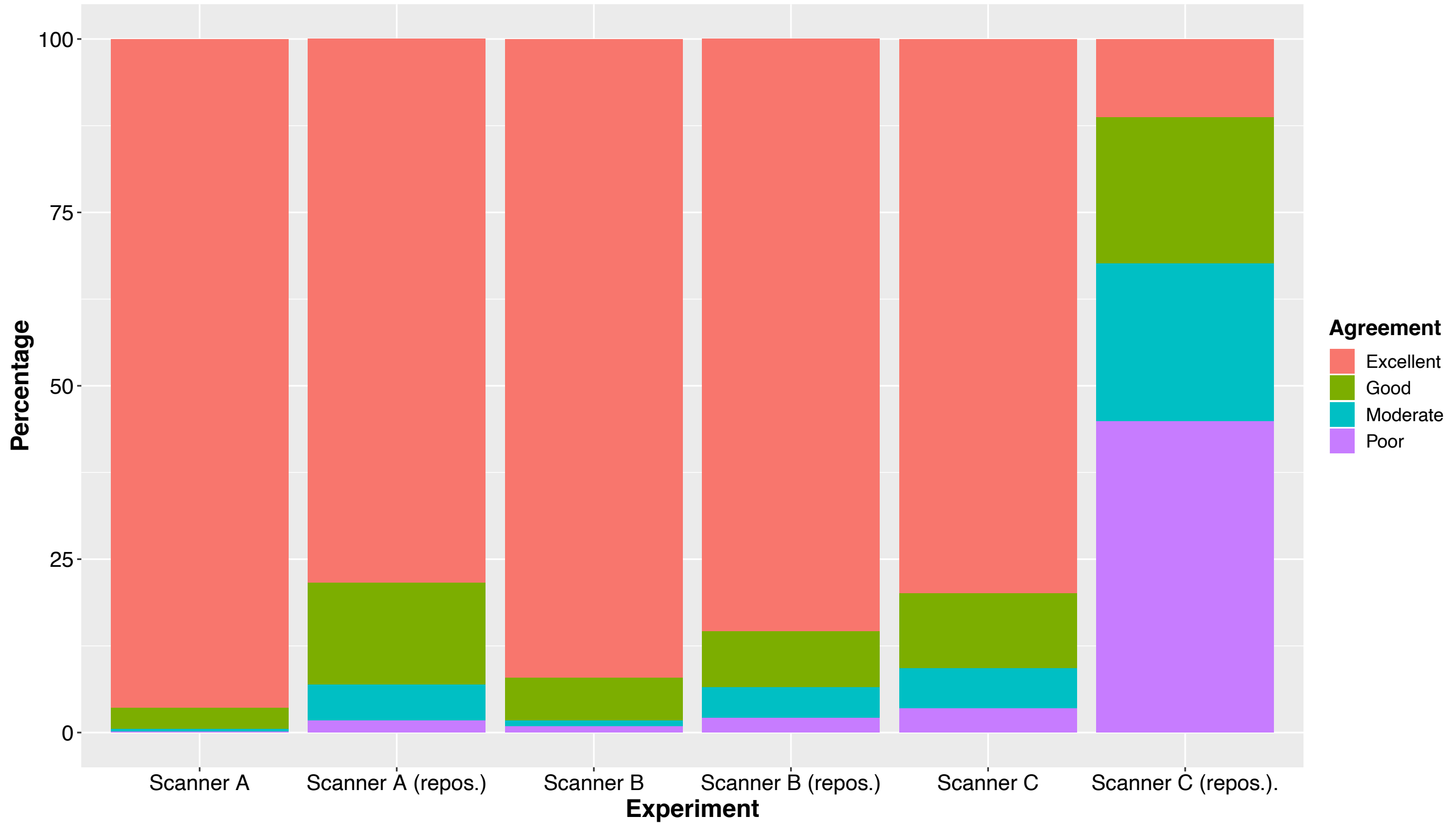

Figure 5. Repeatability of radiomic features. The results are reported for repeatability without phantom repositioning on scanners A, B, and C, and with phantom repositioning (repos.).

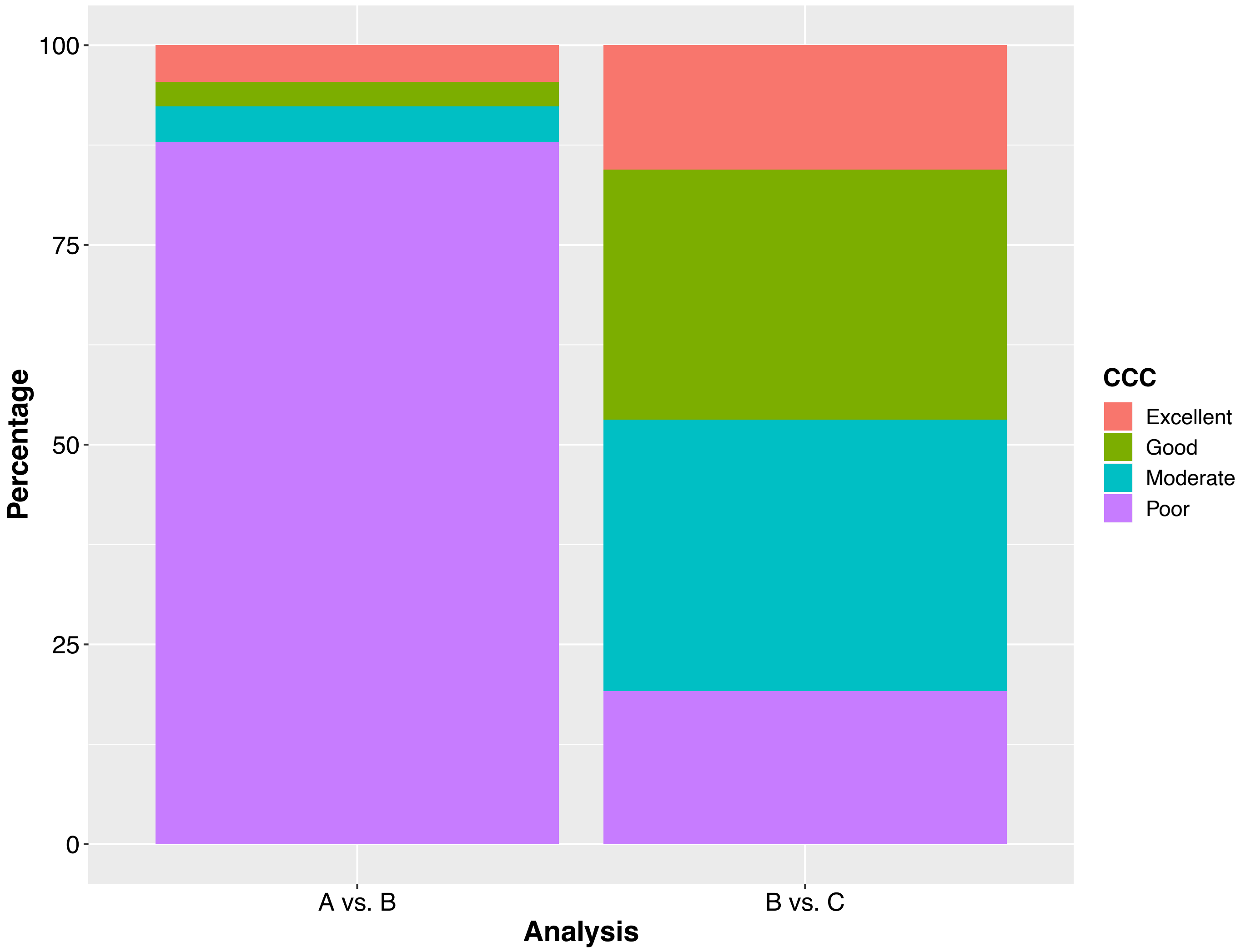

Figure 6. Reproducibility of radiomics features for scanners of equal magnetic field strength (1.5 T) and acquisition parameters, but different manufacturers (A vs. B) and for scanners of the same manufacturer but different magnetic field strengths - 1.5 T and 3 T - (B vs. C).

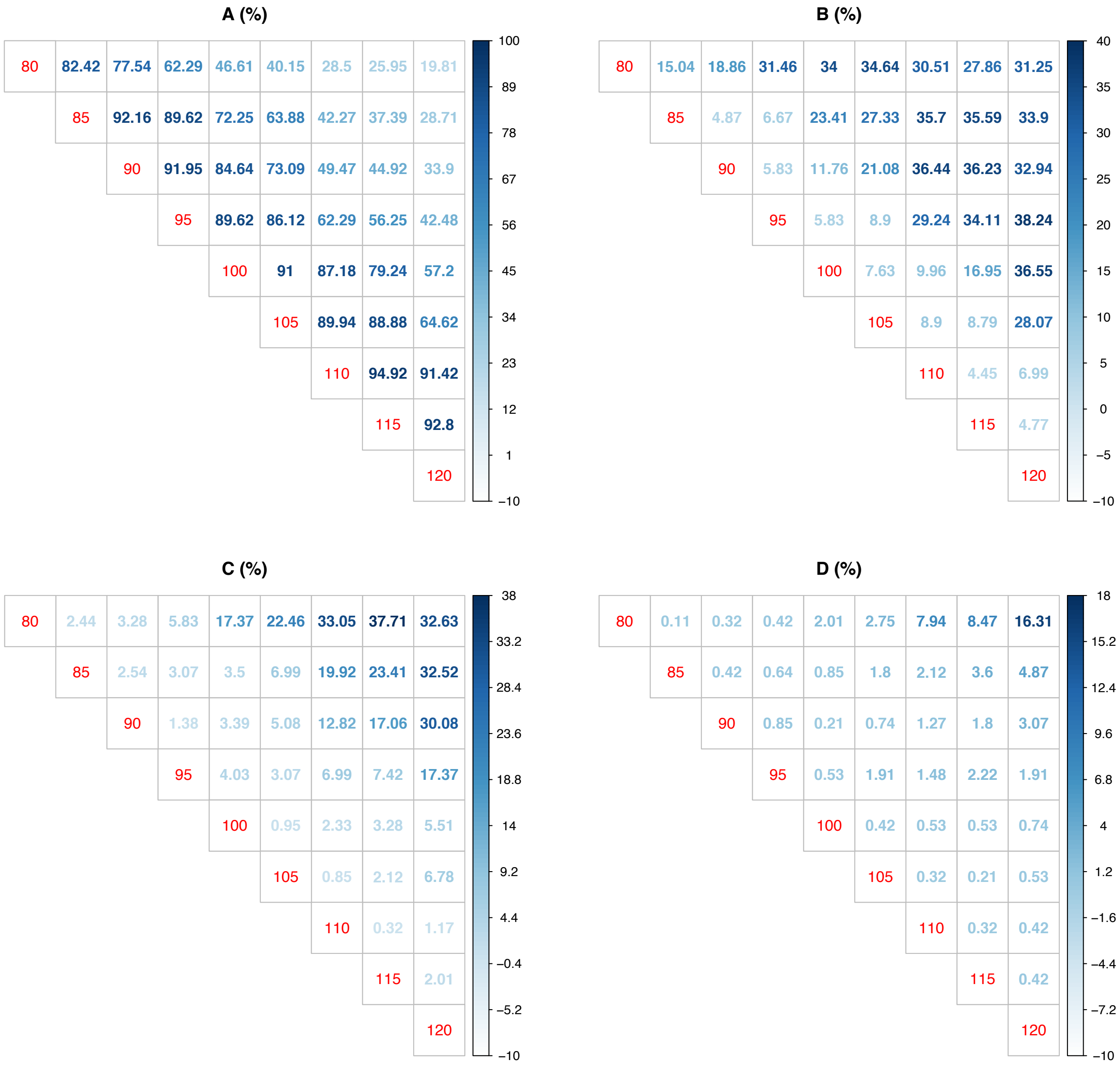

Figure 7. Reproducibility of radiomic features with varying TEs. Reproducibility was assessed for all possible combinations of TEs between 80 ms and 120 ms in 5 ms intervals. (A) Percentage of features with excellent reproducibility; (B) Percentage of features with good reproducibility; (B) Percentage of features with moderate reproducibility; (C) Percentage of features with poor reproducibility.

## Tables

Table 1. Scanners properties and imaging sequence parameters. TR = Repetition Time; TE = Echo Time; CNR – Contrast-to-Noise ratio; FoV = Field of View.

| Scanner | Field strength [T] | Vendor | Experiment | Acquisition | TR [ms] | TE [ms] | CNR | Slice thickness [mm] | Spacing between slices [mm] | Pixel Spacing [mm x mm] | FoV [mm x mm] |
|---|---|---|---|---|---|---|---|---|---|---|---|
| A | 1.5 | GE | Repeatability A/ Reproducibility A vs. B | 2D | 4763 | 109 | 64.90 | 5 | 5.5 | 0.6 x 0.6 | 320 x 320 |
| B | 1.5 | Philips | Reproducibility A vs. B | 2D | 4700 | 110 | 24.71 | 5 | 6 | 0.6 x 0.6 | 320 x 320 |
| | | | Repeatability B 3D | 3D | 1050 | 160 | 455.52 | 1 | 1 | 1 x 1 | 320 x 320 |
| | | | Reproducibility TE/TR | 2D | 5000 | 80 | 24.47 | 5 | 6 | 0.6 x 0.6 | 320 x 320 |
| | | | | | | 85 | 23.92 | | | | |
| | | | | | | 90 | 24.67 | | | | |
| | | | | | | 95 | 22.83 | | | | |
| | | | | | | 100 | 19.16 | | | | |
| | | | | | | 105 | 14.89 | | | | |
| | | | | | | 110 | 12.15 | | | | |
| | | | | | | 115 | 12.51 | | | | |
| | | | | | | 120 | 12.28 | | | | |
| | | | | | 4405 | 100 | 21.84 | | | | |
| | | | Repeatability B 2D/ Reproducibility B vs. C | 2D | 3750 | 90 | 58.86 | 5 | 5 | 0.6 x 0.6 | 340 x 340 |
| C | 3 | Philips | Repeatability C/ Reproducibility B vs. C | 2D | 3750 | 90 | 176.55 | 5 | 5 | 0.6 x 0.6 | 340 x 340 |

Table 2. Classes of stability of radiomic features.

| **Repeatability/Reproducibility** | **ICC or CCC value** |
|---|---|
| Excellent | $ICC\ or\ CCC > 0.9$ |
| Good | $0.75 < ICC\ or\ CCC \leq 0.9$ |
| Moderate | $0.5 < ICC\ or\ CCC \leq 0.75$ |
| Poor | $ICC\ orCCC \leq 0.5$ |

Table 3. List of features showing excellent repeatability with and without phantom repositioning and excellent repeatability across manufacturers and magnetic field strengths (Shape features, expected to be independent of the experiment settings, are not reported in this Table). Each feature is indicated in the form: ImageType_Class_FeatureName.

| |
|---|
| log.sigma.6.mm.3D_firstorder_TotalEnergy |
| exponential_firstorder_Energy |
| exponential_firstorder_TotalEnergy |
| logarithm_glrlm_RunLengthNonUniformity |
| original_glrlm_RunLengthNonUniformity |
| square_firstorder_Energy |
| square_firstorder_TotalEnergy |
| squareroot_glrlm_RunLengthNonUniformity |
| wavelet.HH_firstorder_Energy |
| wavelet.HH_firstorder_TotalEnergy |
| wavelet.HH_ngtdm_Coarseness |
| wavelet.HL_firstorder_Energy |
| wavelet.HL_firstorder_TotalEnergy |
| wavelet.HL_glrlm_RunLengthNonUniformity |
| wavelet.HL_ngtdm_Coarseness |
| wavelet.LH_firstorder_Energy |
| wavelet.LH_firstorder_TotalEnergy |
| wavelet.LH_glrlm_RunLengthNonUniformity |
| wavelet.LH_ngtdm_Coarseness |
| wavelet.LL_glrlm_RunLengthNonUniformity |

Table 4. Repeatable 3D features. List of 3D radiomic features showing excellent repeatability in more than 80% of the image filters.

| Features | Number of Features (%) |
|---|---|
| firstorder_Energy | 14 (100) |
| firstorder_TotalEnergy | 14 (100) |
| glcm_Correlation | 12 (86) |
| glcm_Idn | 12 (86) |
| glcm_Imc1 | 13 (93) |
| glcm_Imc2 | 12 (86) |
| gldm_DependenceNonUniformity | 14 (100) |
| gldm_GrayLevelNonUniformity | 12 (86) |
| glrlm_GrayLevelNonUniformity | 13 (93) |
| glrlm_RunLengthNonUniformity | 13 (93) |
| glszm_GrayLevelNonUniformity | 14 (100) |
| glszm_SizeZoneNonUniformity | 12 (86) |
| ngtdm_Coarseness | 14 (100) |

Table 5. List of texture features showing, across all scanners, high correlation with shape features and uninformative about texture. Each feature is indicated in the form: ImageType_Class_FeatureName.

| | |
|---|---|
| log.sigma.6.mm.3D_firstorder_Energy | log.sigma.6.mm.3D_firstorder_TotalEnergy |
| log.sigma.6.mm.3D_ngtdm_Coarseness | wavelet.LH_firstorder_Energy |
| wavelet.LH_firstorder_TotalEnergy | wavelet.LH_glrlm_RunLengthNonUniformity |
| wavelet.HL_firstorder_Energy | wavelet.HL_firstorder_TotalEnergy |
| wavelet.HL_glrlm_RunLengthNonUniformity | wavelet.HH_firstorder_Energy |
| wavelet.HH_firstorder_TotalEnergy | wavelet.HH_glrlm_RunLengthNonUniformity |
| wavelet.HH_ngtdm_Coarseness | wavelet.LL_glrlm_RunLengthNonUniformity |
| square_firstorder_Energy | square_firstorder_TotalEnergy |
| square_glszm_GrayLevelNonUniformity | exponential_firstorder_Energy |
| exponential_firstorder_TotalEnergy | |

# Supplementary material

## S.1 SNR calculation

Signal-to-noise ratio (SNR) was calculated on z-score normalized images using equation (s1) where the mean signal intensity of A, $\mu_{SI_A}$, was measured in the green ROI in Figure 3 and the standard deviation of the signal intensity of background air, $\sigma_{SI_{air}}$, was measured in the red ROI.

$$SNR = \frac{\mu_{SI_A}}{\sigma_{SI_{air}}} \qquad \text{(s1)}$$

## S.2 Additional sequences information table

In Table S1 additional scanners properties and imaging parameters are listed.

## S.3 Feature extraction settings

In Table S2 the settings contained in the parameter file for the radiomic features extraction with the package PyRadiomics are reported.

## S.4 Feature extraction settings

In Table S3 the list of extracted features by feature category is reported.

## S.5 Correlation matrix (shape and non-shape features)

The correlation between shape and non-shape features was determined using the Spearman correlation method. This was calculated for both original and filtered images. In Figure S1, it is shown the correlation of the different non-shape features with the shape features for the original images acquired on scanner A.

It is possible to observe that non-shape features can be correlated with other shape features besides volume, and this is the main reason for the consideration of all shape features in this analysis.

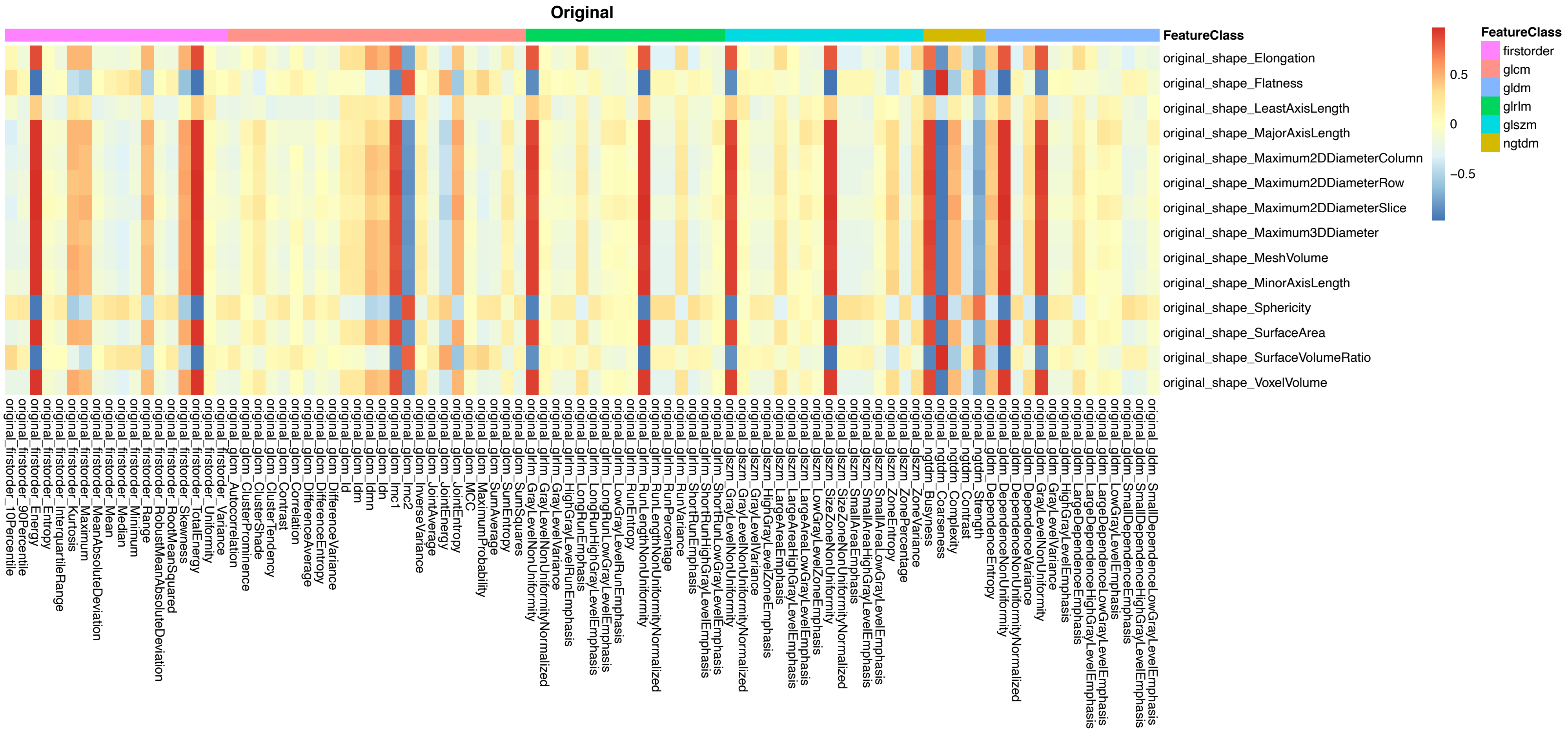

Figure S1. **(Supplementary material)** Correlation matrix of shape and non-shape radiomic features for the original images acquired on scanner A.

Table S1. **(Supplementary material)** Additional scanners properties and imaging sequence parameters. TR = Repetition Time; TE = Echo Time; SNR – Signal-to-Noise ratio.

| Scanner | Field strength [T] | Vendor | Experiment | Acquisition | TR [ms] | TE [ms] | SNR | Number of Averages | Total scan time |
|---|---|---|---|---|---|---|---|---|---|
| A$^*$ | 1.5 | GE | Repeatability A/ Reproducibility A vs. B | 2D | 4763 | 109 | 89.68 | 1 | 3 min 32 s |
| B$^+$ | 1.5 | Philips | Reproducibility A vs. B | 2D | 4700 | 110 | 47.42 | 1 | 5 min 10 s |
| | | | Repeatability B 3D | 3D | 1050 | 160 | 12.50 | 2 | 6 min 47 s |
| | | | Reproducibility TE/TR | 2D | 5000 | 80 | 16.65 | 1 | 4 min 30 s |
| | | | | | | 85 | 19.29 | | |
| | | | | | | 90 | 21.64 | | |
| | | | | | | 95 | 19.83 | | |
| | | | | | | 100 | 24.16 | | |
| | | | | | | 105 | 22.09 | | |
| | | | | | | 110 | 23.04 | | |
| | | | | | | 115 | 25.46 | | |
| | | | | | | 120 | 23.99 | | |
| | | | | | 4405 | 100 | 21.28 | 1 | 3 min 58 s |
| | | | Repeatability B 2D/ Reproducibility B vs. C | 2D | 3750 | 90 | 12.00 | 1 | 3 min 30 s |
| C$^+$ | 3 | Philips | Repeatability C/ Reproducibility B vs. C | 2D | 3750 | 90 | 18.21 | 1 | 5 min 15 s |

* - On scanner A, a 24 channels anterior body array coil and a 24 channels spine array coil were used; + - On scanners B and C, a 32 channels torso array coil was used.

Table S2. **(Supplementary material)** Settings in the parameter files used for the radiomic features extraction with the package PyRadiomics.

| **Setting** | **Value** |
|---|---|
| normalize | true |
| normalizeScale | 100 |
| preCrop | true |
| force2D | true (2D extraction)<br>false (3D extraction) |
| force2Ddimension | 0 (2D extraction only) |
| geometryTolerance | 1.e+4 |
| binwidth | 10 (scanner A)<br>5 (scanners B and C) |
| voxelArrayShift | 300 |
| imageType | Original: {}<br>LoG: {'sigma': [6]}<br>Wavelet: {'level': 2}<br>Square: {}<br>SquareRoot: {}<br>Logarithm: {}<br>Exponential: {} |
| featureClass | Shape:<br>firstorder:<br>glcm:<br>glrlm:<br>glszm:<br>ngtdm:<br>gldm: |

Table S3. **(Supplementary material)** List of radiomic features per category extracted with the package PyRadiomics. Shape features are extracted from the original images only, whilst the features of the other categories are extracted from both the original and filtered images. The definition of the features listed in this table is available at https://pyradiomics.readthedocs.io/en/latest/features.html

| Shape | First Order | GLCM | GLRLM |
|---|---|---|---|
| Elongation | 10Percentile | Autocorrelation | GrayLevelNonUniformity |
| Flatness | 90Percentile | ClusterProminence | GrayLevelNonUniformityNormalized |
| LeastAxisLength | Energy | ClusterShade | GrayLevelVariance |
| MajorAxisLength | Entropy | ClusterTendency | HighGrayLevelRunEmphasis |
| Maximum2DDiameterColumn | InterquartileRange | Contrast | LongRunEmphasis |
| Maximum2DDiameterRow | Kurtosis | Correlation | LongRunHighGrayLevelEmphasis |
| Maximum2DDiameterSlice | Maximum | DifferenceAverage | LongRunLowGrayLevelEmphasis |
| Maximum3DDiameter | MeanAbsoluteDeviation | DifferenceEntropy | LowGrayLevelRunEmphasis |
| MeshVolume | Mean | DifferenceVariance | RunEntropy |
| MinorAxisLength | Median | Id | RunLengthNonUniformity |
| Sphericity | Minimum | Idm | RunLengthNonUniformityNormalized |
| SurfaceArea | Range | Idmn | RunPercentage |
| SurfaceVolumeRatio | RobustMeanAbsoluteDeviation | Idn | RunVariance |
| VoxelVolume | RootMeanSquared | Imc1 | ShortRunEmphasis |
| | Skewness | Imc2 | ShortRunHighGrayLevelEmphasis |
| | TotalEnergy | InverseVariance | ShortRunLowGrayLevelEmphasis |
| | Uniformity | JointAverage | |
| | Variance | JointEnergy | |
| | | JointEntropy | |
| | | MCC | |
| | | MaximumProbability | |
| | | SumAverage | |
| | | SumEntropy | |
| | | SumSquares | |

Table S3 (continued). **(Supplementary material)**

| GLSZM | NGTDM | GLDM |
|---|---|---|
| GrayLevelNonUniformity | Busyness | DependenceEntropy |
| GrayLevelNonUniformityNormalized | Coarseness | DependenceNonUniformity |
| GrayLevelVariance | Complexity | DependenceNonUniformityNormalized |
| HighGrayLevelZoneEmphasis | Contrast | DependenceVariance |
| LargeAreaEmphasis | Strength | GrayLevelNonUniformity |
| LargeAreaHighGrayLevelEmphasis | | GrayLevelVariance |
| LargeAreaLowGrayLevelEmphasis | | HighGrayLevelEmphasis |
| LowGrayLevelZoneEmphasis | | LargeDependenceEmphasis |
| SizeZoneNonUniformity | | LargeDependenceHighGrayLevelEmphasis |
| SizeZoneNonUniformityNormalized | | LargeDependenceLowGrayLevelEmphasis |
| SmallAreaEmphasis | | LowGrayLevelEmphasis |
| SmallAreaHighGrayLevelEmphasis | | SmallDependenceEmphasis |
| SmallAreaLowGrayLevelEmphasis | | SmallDependenceHighGrayLevelEmphasis |
| ZoneEntropy | | SmallDependenceLowGrayLevelEmphasis |
| ZonePercentage | | |
| ZoneVariance | | |